\documentclass[pre,superscriptaddress,twocolumn,nofootinbib,showpacs,floatfix]{revtex4}

\usepackage{amsmath,amsfonts,amssymb}
\usepackage{graphicx}
\usepackage{hyperref}
\usepackage{verbatim}

\newcommand{\be}{\begin{equation}}
\newcommand{\ee}{\end{equation}}
\newcommand{\bse}{\begin{subequations}}
\newcommand{\ese}{\end{subequations}}
\newcommand{\eq}[1]{Eq.~(\ref{#1})}

\newcommand{\RR}{\mathbb{R}}

\newcommand{\br}{R}

\newcommand{\bphi}{\Phi}
\newcommand{\bx}{X}
\newcommand{\by}{Y}
\newcommand{\bz}{Z}
\newcommand{\bp}{P}
\newcommand{\bpsi}{\Psi}
\newcommand{\bn}{N}

\begin{document}

\title{Breaking the Rules for Topological Defects: Smectic Order on Conical Substrates}

\author{Ricardo A. Mosna}
\affiliation{Department of Physics and Astronomy, University of Pennsylvania, Philadelphia, PA 19104-6396, USA}
\affiliation{Departamento de Matem\'atica Aplicada, Universidade Estadual de Campinas, 13083-859, Campinas, SP, Brazil}
\author{Daniel A. Beller}  
\affiliation{Department of Physics and Astronomy, University of Pennsylvania, Philadelphia, PA 19104-6396, USA}
\author{Randall D. Kamien}  
\affiliation{Department of Physics and Astronomy, University of Pennsylvania, Philadelphia, PA 19104-6396, USA}

\date{\today}

\begin{abstract}
Ordered phases on curved substrates experience a complex interplay of ordering and intrinsic curvature, commonly producing frustration and singularities. This is an especially important issue in crystals as ever-smaller scale materials are grown on real surfaces; eventually, surface imperfections are on the same scale as the lattice constant. Here, we gain insights into this general problem by studying two-dimensional smectic order on substrates with highly localized intrinsic curvature, constructed from cones and their intersections with planes.  In doing so we take advantage of fully tractable ``paper and tape'' constructions, allowing us to understand, in detail, the induced cusps and singularities.
\end{abstract}

\maketitle

\section{Introduction and Summary}
Dramatic progress has been made in algorithmic origami; it is now possible to design nearly arbitrary three-dimensional constructions out of unstretched \cite{lang} or {\sl nearly} unstretched \cite{demaine} plaquettes, isometric to pieces of the Euclidean two-plane.  At the same time, there has been theoretical and technological interest in crystalline and liquid crystalline order on curved substrates \cite{xing,bowi2000,vitelucks06,jia09,park92}.  The latter problem can be studied in reduced complexity by considering surfaces with vanishing Gaussian curvature except at isolated points and curves. Were we to consider only intrinsic interactions between the substrate and the ordered phase, it follows that the in-plane positional and orientational order would be completely determined by the folds and conical points of the substrate.  The connection between smectic textures, geometric optics \cite{columns,xing}, and shocks \cite{AKS} on flat and curved surfaces makes studies of the smectic phase amenable to exact analytic study \cite{BPS,AKM} while capturing the salient features of broken translational and rotational invariance \cite{Chen}.  Combined with the simplified geometries we consider, we are led to highly tractable models of order on curved backgrounds.

\begin{figure}[h]
\begin{center}
\includegraphics[width=0.22\textwidth]{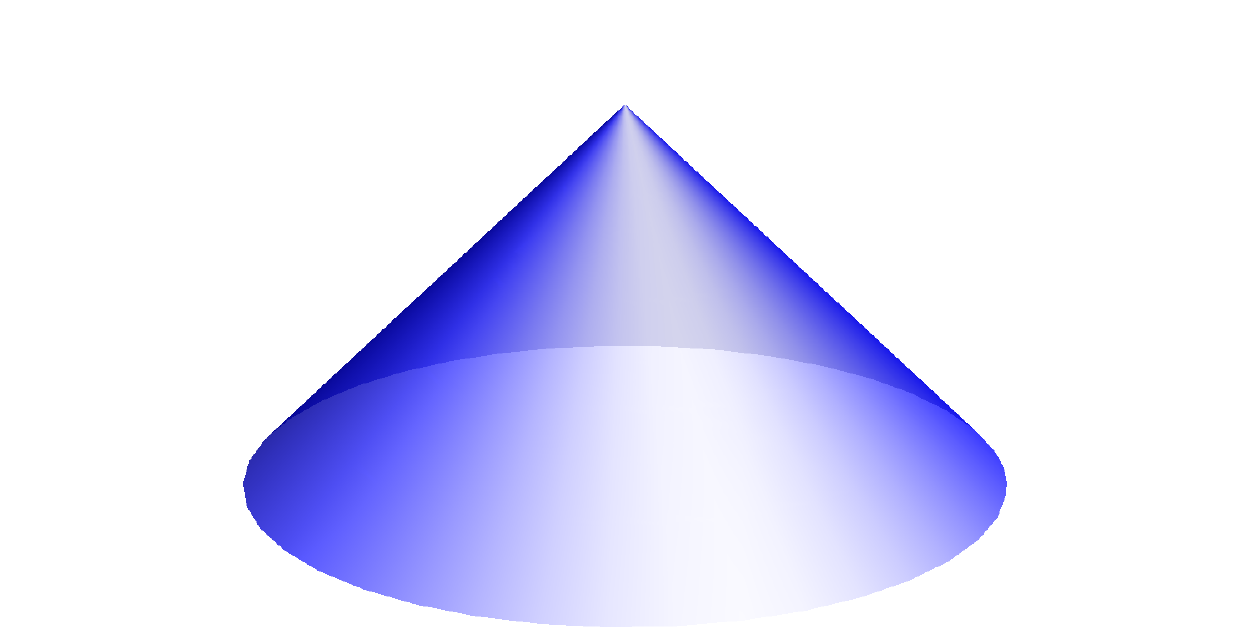}
\includegraphics[width=0.22\textwidth]{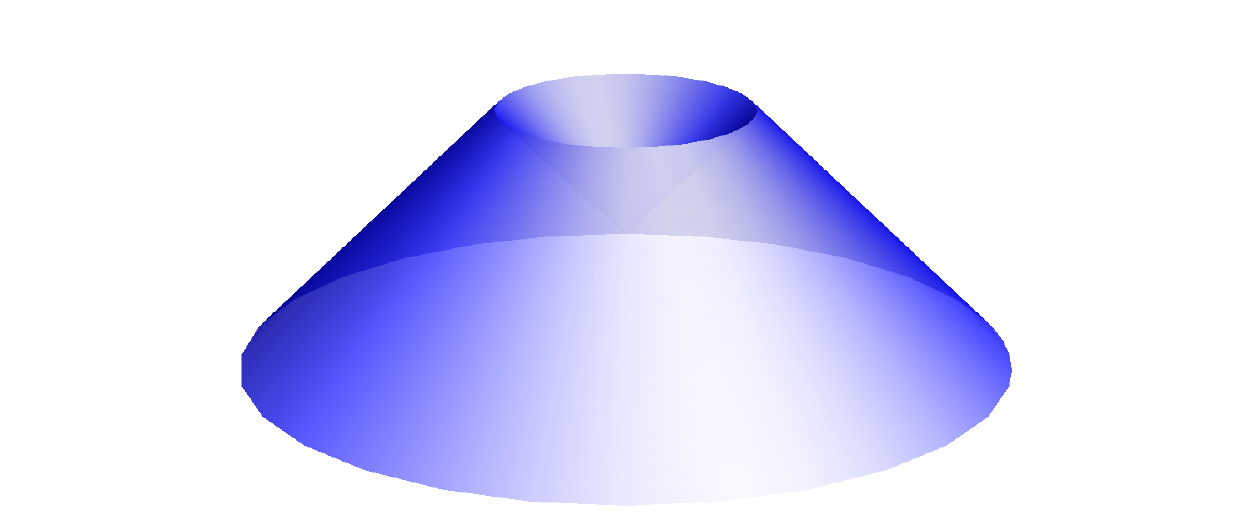}
\includegraphics[width=0.22\textwidth]{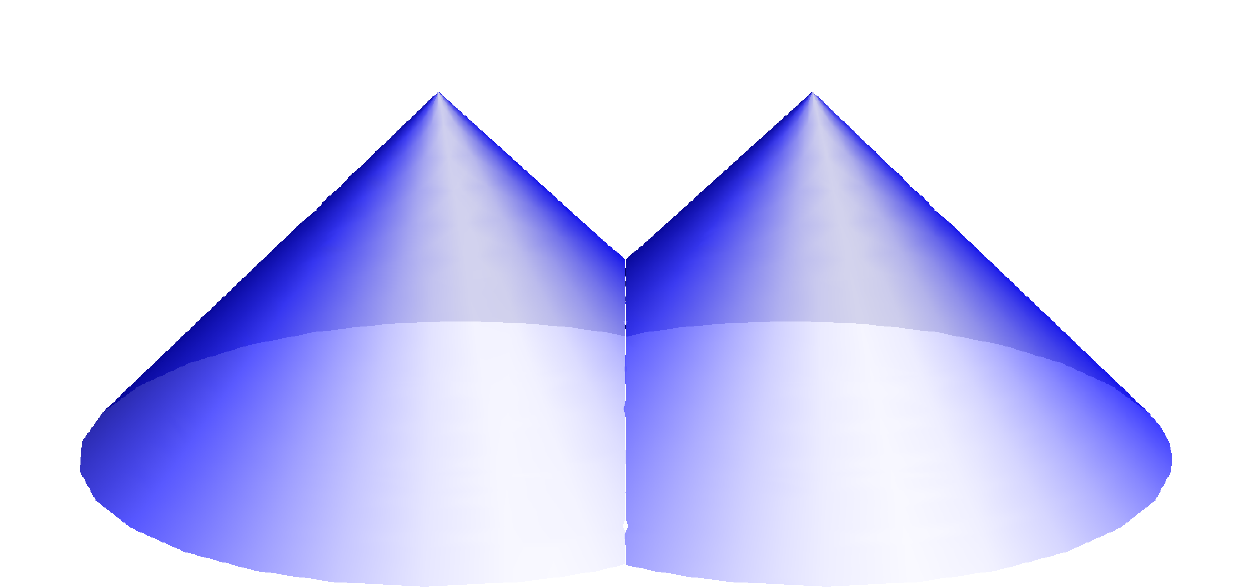}
\includegraphics[width=0.22\textwidth]{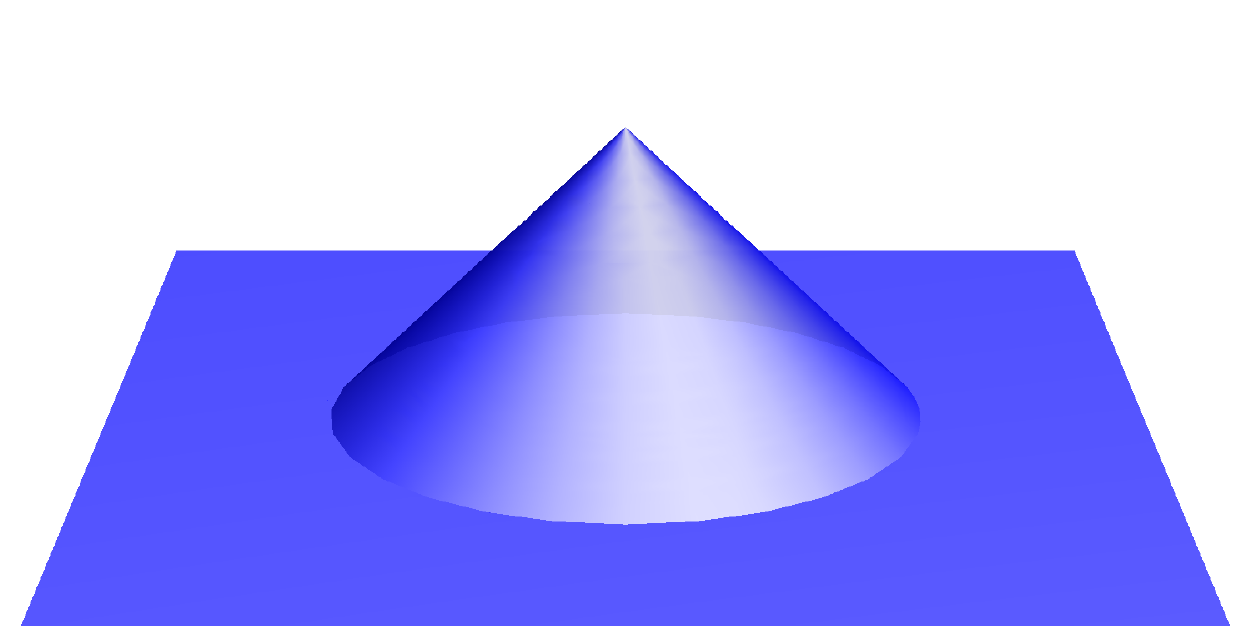}
\end{center}
\caption{Substrates with conical defects. Notice that all of these surfaces are intrinsically flat almost everywhere -- they fail to be flat at points and lines only.}
\label{fig:substrates}
\end{figure}

Here we study equally-spaced smectic textures on infinite cones, cones interesecting with planes, and cones intersecting with cones, the latter two standing in for simple bumps on surfaces and saddle-like regions, respectively; see Fig.~\ref{fig:substrates}.   Not only do we see the occurrence of focal lines and cusps in the ensuing smectic textures, but we also see violations of the rules that usually govern the schlieren textures of the sample.  Recall that curved geometry interacts with topological defects \cite{bausch} to alter the conservation of topological charge in much the same way the background intrinsic curvature changes the rules for the sum of the interior angles of a triangle.  In the case of schlieren textures in flat geometries, defects mark the confluence of an even number of dark brushes.  However, even this simple counting rule is violated on curved surfaces, as we will demonstrate.  

\begin{figure}[h]
\includegraphics[width=0.42\textwidth]{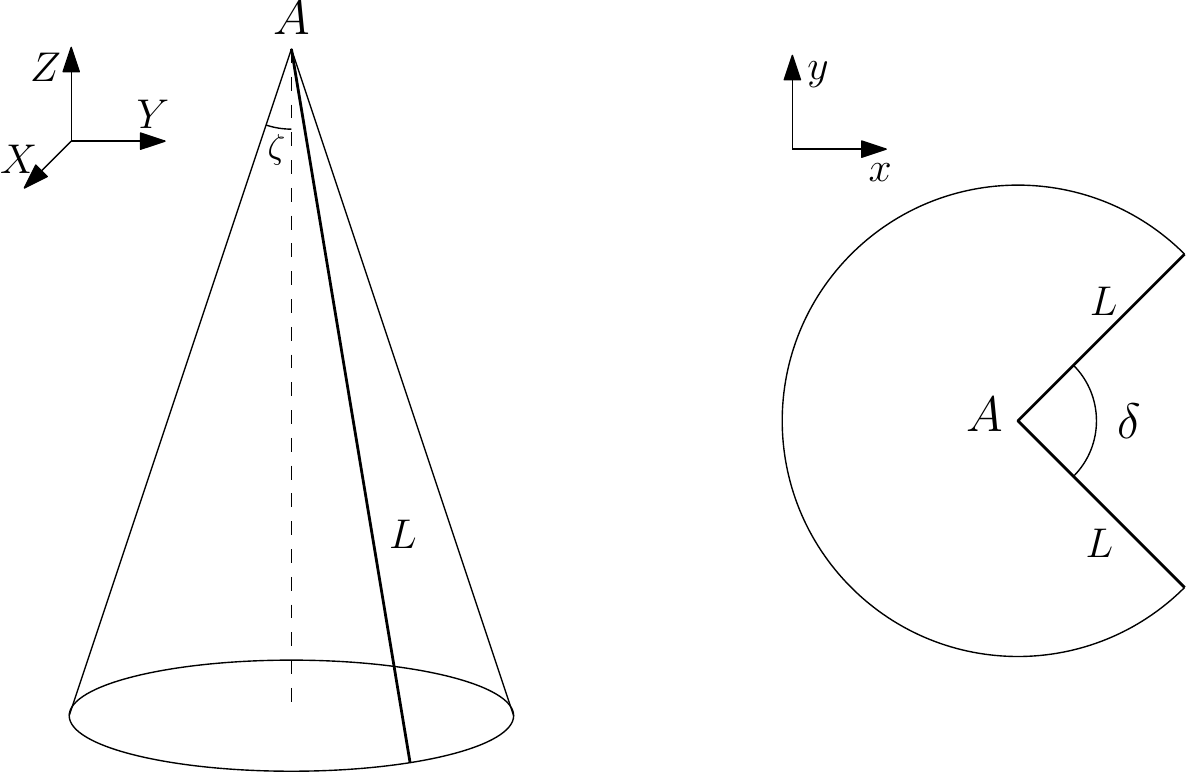}
\caption{A cone is isometric to a cut planar disk with two radial lines (denoted by $L$ above) identified. The point $A$ is the apex of the cone.}
\label{fig:flattened_cone}
\end{figure}

We commence with a cone $C$ embedded in $\RR^3$.  $C$ is a singular surface which is flat everywhere except at its apex, where all the Gaussian curvature is concentrated. The geometry of $C$ may be conveniently examined by cutting the cone along a radial line $L$ and laying it flat on a plane (Fig.~\ref{fig:flattened_cone}). This way, $C$ looks like a disk with a circular sector of angle $\delta$ removed and with its two straight edges identified. The angle $\delta$ is called the deficit angle. A direct application of the Gauss-Bonnet Theorem shows that $\delta$ is also the total Gaussian curvature of any region of the cone which contains its apex.

In order to establish notation, let $\bx,\by,\bz$ and $x,y$ be Cartesian coordinates as in Fig.~\ref{fig:flattened_cone}. A parametrization of $C$ is given by:
\begin{align*}
\bx(\br,\bphi) =& \br \cos\bphi \sin\zeta, \\
\by(\br,\bphi) =& \br \sin\bphi \sin\zeta, \\
\bz(\br,\bphi) =& -\br \cos\zeta,
\end{align*}
where, as can be easily seen, the apex angle $2\zeta$ is related to $\delta$ by $\sin\zeta=\frac{2\pi-\delta}{2\pi}$.
In terms of $x,y$ (see Fig.~\ref{fig:flattened_cone}) and their polar coordinates, $r =\sqrt{x^2+y^2}$ and $\tan\phi=y/x$, we have 
\begin{eqnarray}\label{eq:transform}
\br &=&  r,\nonumber\\
\bphi &=& \left(\phi-\frac{\delta}{2}\right) \csc\zeta.
\end{eqnarray}

\section{Building the Layers}

\begin{figure}[h]
\includegraphics[width=0.3\textwidth]{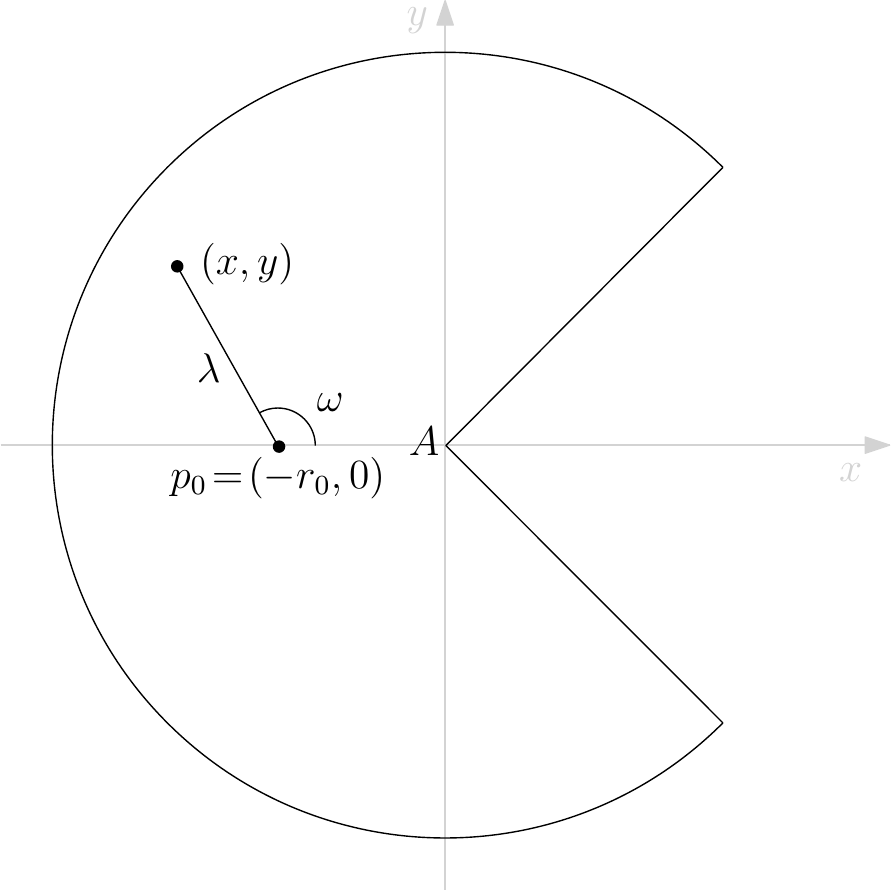}
\caption{Cut-disk view of the cone, showing a parametrization of geodesics emanating from a point disclination.}
\label{fig:parametrization}
\end{figure}

A necessary condition for the layers to be uniformly spaced is that their normal vector field points along geodesics of the surface~\cite{Kleman,columns}. In the flattened model, these geodesics are just the straight lines of the plane. We know that defects (even in flat space) tend to concentrate on lower dimensional sets in order to save energy so, in a 2-dimensional surface, this means that point defects are favored and this gives rise to layered structures in the form of wavefronts emanating from a point. Note that the case where this point is taken to infinity formally corresponds to a defect free configuration. We are thus led to consider a  wavefront starting at some point $\bp_0$, whose corresponding point $p_{0}$ on the cut disk lies at a distance $r_0$ from the disk center, which maps to the cone apex. Experimentally, this scenario can be created deliberately with a colloidal particle that induces homeotropic anchoring for the molecules of the liquid crystal. We can always cut the cone so that $L$ is exactly opposed to $\bp_0$ and then choose coordinates such that $L$ lies in the $XZ $ plane. By doing so, we have $p_0=(-r_0,0)$. The geodesic ``light rays" may then be parametrized by $x(\lambda)=-r_0+\lambda\cos\omega$, $y(\lambda)=\lambda\sin\omega$; see Fig.~\ref{fig:parametrization}. Whereas computing the geodesics on a generic surface is nontrivial, for a conical substrate we have a simple analytic mapping of straight lines on the cut disk to geodesics on the cone. The associated smectic layers are concentric circles centered at $p_0$.  Note that we could choose the cut $L$ along any direction we like as long as we identify the two edges.  Were we to do so, when a straight line in the {\sl flattened} model hits one of these cuts, we would continue it with a straight line emanating from the other cut, making the same angle with the new edge.  This ensures that the geodesics remain straight and demonstrates that the smectic texture is independent of the choice of $L$.  The presence of any cusps or grain boundaries in the smectic does not result from the flattened geometry -- all this could be computed directly on the cone, for instance.  Note that we could, alternatively, construct geodesics on a (full) two-disk, parameterized by $\bx$ and $\by$ with induced metric:
\begin{widetext}
\begin{equation}\label{eq:metric}
ds^2 =\left[\left(1+{\cot^2\zeta}\frac{\bx^2}{\bx^2+\by^2}\right) d\bx^2 + {\cot^2\zeta}\frac{2\bx\by}{\bx^2+\by^2}d\bx d\by + \left(1+{\cot^2\zeta}\frac{\by^2}{\bx^2+\by^2}\right)d\by^2\right] .
\end{equation}
\end{widetext}

Let $\bx(\lambda),\by(\lambda),\bz(\lambda)$ be the coordinates on the cone of the geodesics defined above. A straightforward calculation shows that
\begin{equation}
\left[\begin{matrix}
\dot{\bx}(\lambda) \\ \dot{\by}(\lambda) \\ \dot{\bz}(\lambda)
\end{matrix}\right]=k
\left[\begin{matrix}\cos\bpsi \, \bx+\sin\bpsi \, \by \\ -\sin\bpsi \, \bx+\cos\bpsi \, \by \\ \cos\bpsi\bz\end{matrix}\right],
\end{equation}
where 
\begin{eqnarray}
k&=&\frac{\sin^2\zeta}{\bx^2+\by^2} \sqrt{(\lambda-r_0\cos\omega)^2+\left(\frac{r_0\sin\omega}{\sin\zeta}\right)^2},\\
\bpsi&=&\arctan\left(\frac{r_0 \sin\omega}{(\lambda-r_0\cos\omega)\sin\zeta}\right).
\end{eqnarray}
The unit vector field obtained after normalizing this expression is given by 
\begin{equation}\label{tang_vector}
\bn=\frac{1}{\sqrt{1+\cot^2\zeta\cos^2\bpsi}}\left[\begin{matrix} \cos(\bphi-\bpsi) \\ \sin(\bphi-\bpsi) \\ -\cos\bpsi \cot\zeta \end{matrix}\right].
\end{equation}
Therefore, the projection of $\bn$ onto the $\bx\by$ plane makes an oriented angle $\bphi-\bpsi$ with the $\bx$ axis (note that $\bpsi$ depends on $\bx$ and $\by$ through $\lambda$ and $\omega$ (Fig.~\ref{fig:parametrization})).  The corresponding vector in the $\bx\by$ plane points along the unit direction
\begin{equation}\label{proj_tang_vector}
\bn_p =\left[\begin{matrix} \cos(\bphi-\bpsi) \\ \sin(\bphi-\bpsi) \end{matrix}\right].
\end{equation}
The projected layers can also be directly obtained in these coordinates by drawing lines which are everywhere perpendicular to $\bn_p$, with respect to the induced cone metric (\ref{eq:metric}).
Thus, a single prescribed defect, together with the constraint of equal layer spacing, uniquely determines the layer structure everywhere.

\begin{figure}[h]
\begin{center}
\includegraphics[width=0.49\textwidth]{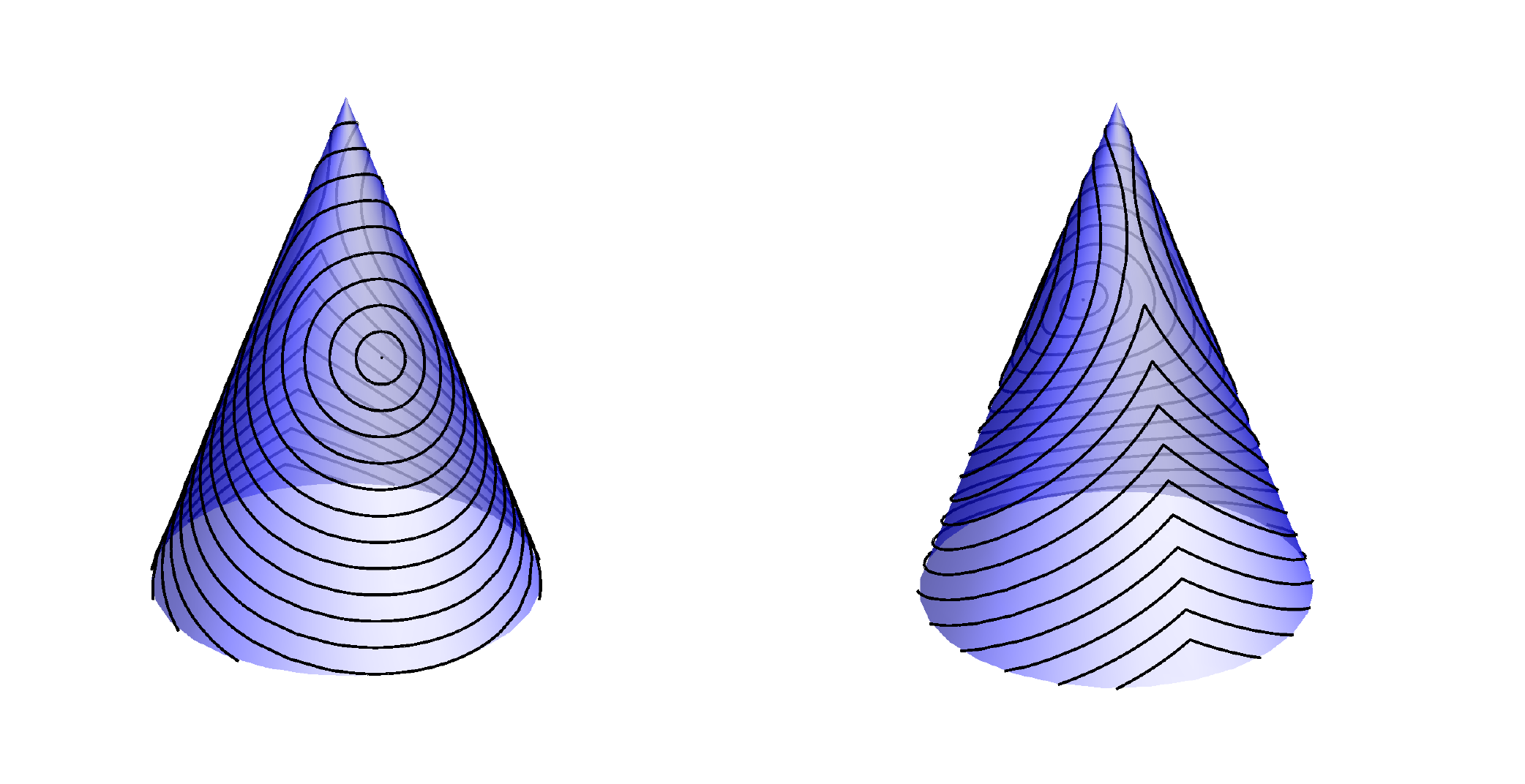}
\end{center}
\caption{Layer structure on the 3D cone for a deficit angle $\delta=5\pi/4$. The image on the right shows the back of the image on the left and {\sl vice versa}.}
\label{fig:cone3d}
\end{figure}

\begin{figure}[h]
\begin{center}
\includegraphics[width=0.49\textwidth]{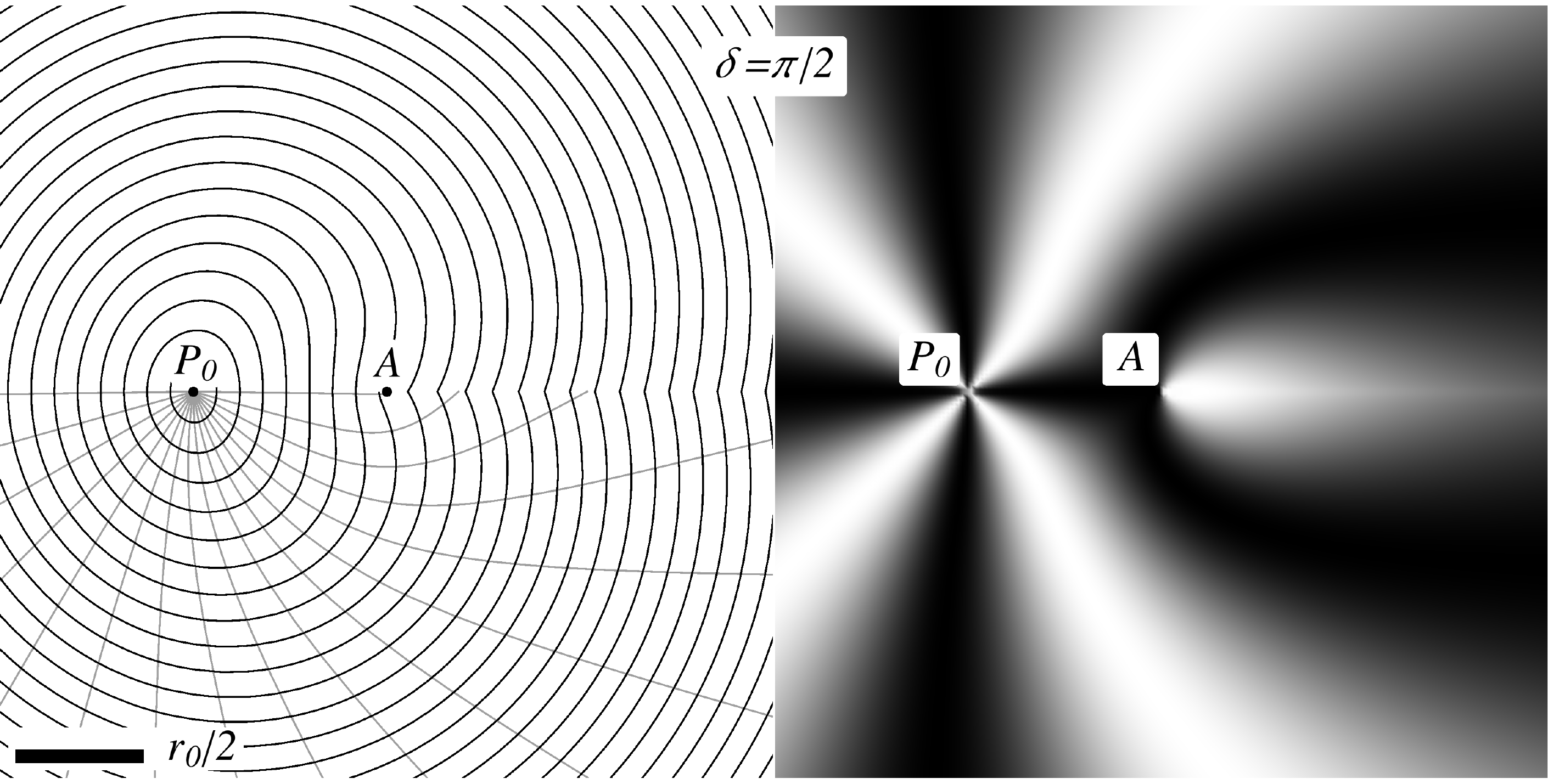} 
\includegraphics[width=0.49\textwidth]{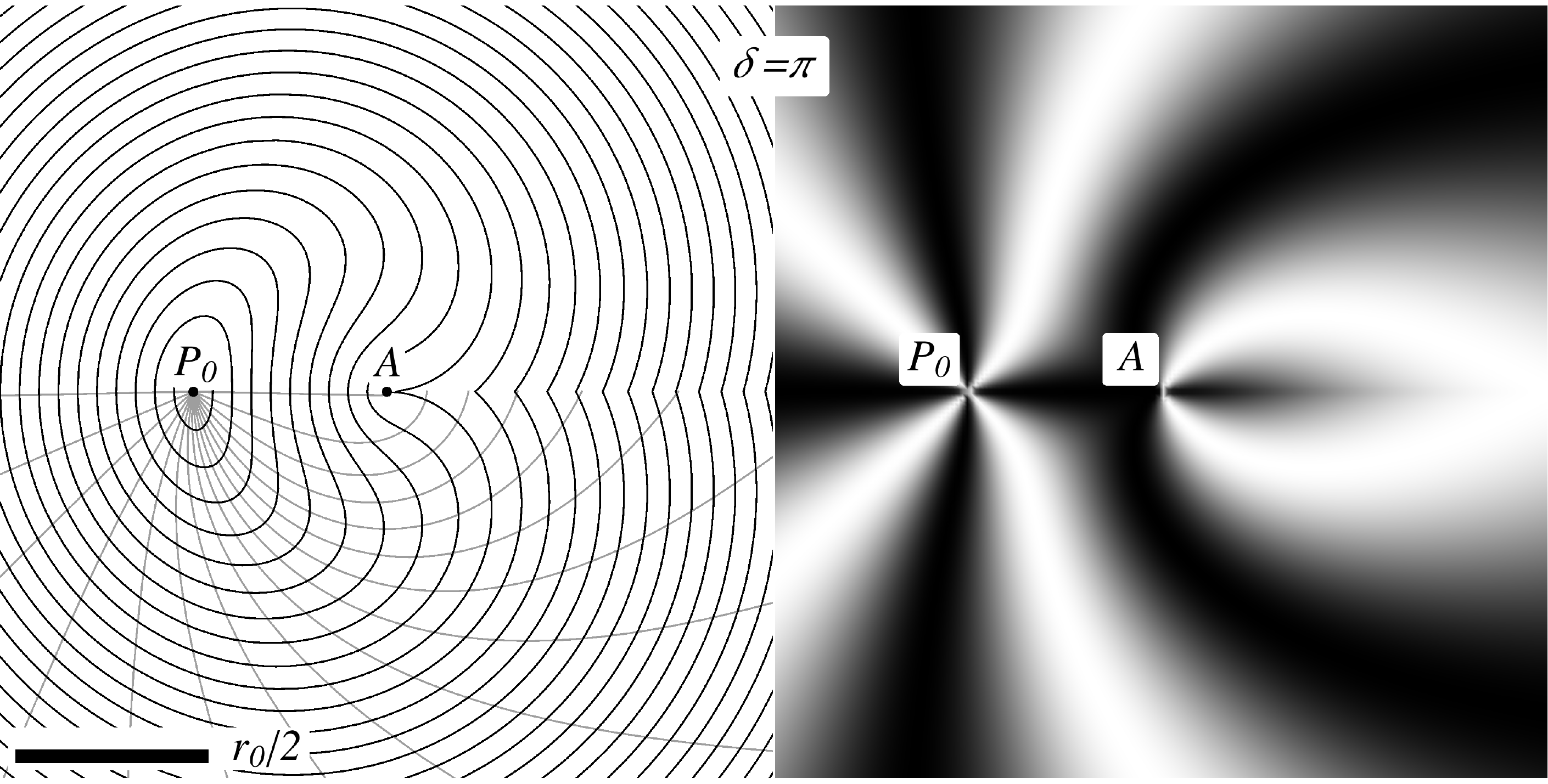} 
\includegraphics[width=0.49\textwidth]{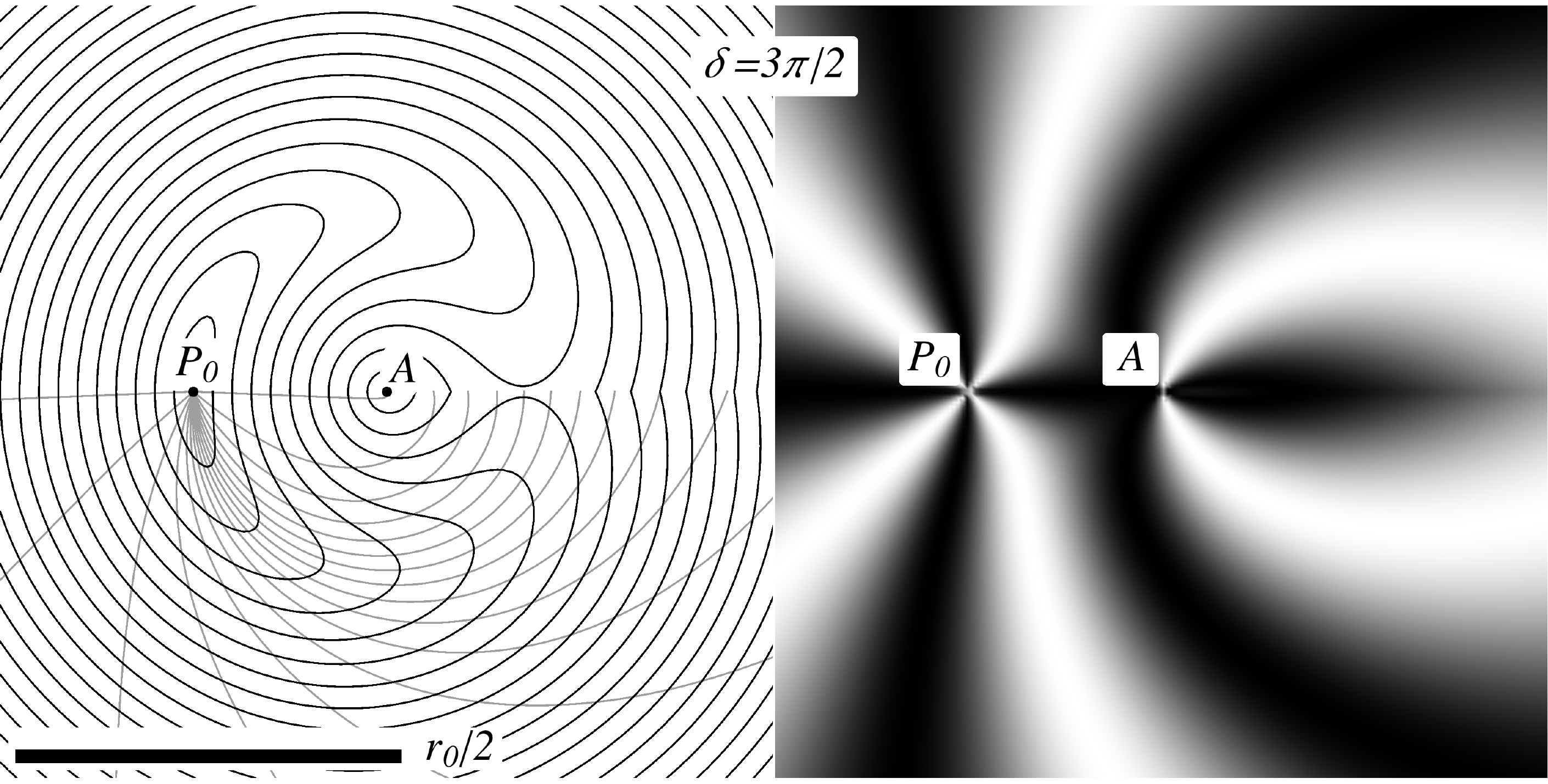} 
\end{center}
\caption{The left panels show the layer structure (black lines) superimposed on geodesics (gray lines on the bottom half) on the cone, as seen from above (i.e., projected on the $\bx\by$ plane), for deficit angles $\delta=\pi/2,\pi$, and $3\pi/2$, respectively. The right panels show the corresponding schlieren textures for the same deficit angles. $A$ and $P_0$ label the cone apex and the disclination location, respectively.}
\label{fig:layers_cone}
\end{figure}

Alternatively, we can compute the layers as the level sets of a function $D(\bp)$ that measures the distance from a given point $\bp$ to the wavefront source $\bp_0$. If the cut line $L$ is appropriately chosen (so that it contains $\bp_0$, for example), the distance between two points $\bp=(\bx,\by,\bz)$ and $\bp_0=(\bx_0,\by_0,\bz_0)$ on the cone is just the planar distance between their counterparts $p=(x,y)$ and $p_0=(x_0,y_0)$ on the cut disk. In terms of the coordinates $R,\bphi$ (see Eq.~(\ref{eq:transform})), this yields
\begin{eqnarray}
D(\bp)&=&\sqrt{(x+r_0)^2 + y^2}\\
&=&\sqrt{R^2+r_0^2+
    2r_0R\cos\left(\frac{\delta}{2}+\bphi\sin{\zeta}\right)}.\nonumber
\end{eqnarray}
The layer structure on the cone, obtained from equally spaced level sets of $D$, is shown in Fig.~\ref{fig:cone3d}. The projected layers and geodesics seen from above are shown in the left panels of Fig~\ref{fig:layers_cone}. Notice that the positive $\bx$ axis develops a grain boundary for any nontrivial deficit angle.

Indeed, the existence of a cusp is a consequence of the Gauss-Bonnet Theorem \cite{columns}; the maximum cusp angle of the layers at the grain boundary is equal to $\pi-\delta$ and occurs at the cone apex. To see this, consider a closed path consisting of the geodesic at $\omega=\omega_{0}$ from $P_{0}$ to an arbitrary point $P_1$ on the grain boundary, followed by the ``mirror'' geodesic at $\omega=-\omega_{0}$ from $P_1$ back to $P_{0}$. The geodesics form an interior angle $\pi-\alpha_{c}$ on the cone at $P_1$, where $\alpha_{c}$ is the cusp angle formed by the layer at $P_1$. At $P_{0}$, $2\omega_0$ is the interior angle formed by the geodesics on the cone. Because the geodesic curvature is zero on this path, and the integrated Gaussian curvature is simply $\delta$, the Gauss-Bonnet Theorem implies
\begin{equation}  
\delta=2\pi-(\pi-\alpha_{c})-(\pi-2\omega_0)=\alpha_{c}+2\omega_0
\end{equation}
(notice that this equation also follows at once from the flattened model by elementary geometry).
The cusp angle $\alpha_{c}$ is therefore maximized when we take $\omega_0\rightarrow0^{+}$, which corresponds to taking $P_1$ arbitrarily close to the cone apex.
For values of $\delta$ greater than $\pi$, the cusp angle becomes $\pi$ at some point to the right of the cone apex, meaning that the layers turn back toward the apex. Consequently, when $\delta>\pi$ the grain boundary is interrupted by two new point defects: a $+1$-index disclination at the apex and a $-1$-index disclination on the positive $X$-axis! This is shown in the left panels of Fig.~\ref{fig:layers_cone} and also in Fig.~\ref{fig:cone3d}. The negative-index defect results from the fact that the normal of some layers turns through an angle greater than or equal to $\pi$, and the outermost such layer has a self-intersection on the grain boundary, resulting in a locally hyperbolic configuration. By setting $Y=0$, $N_{p} = (0,\pm1)$ and solving for $X$, we discover that the $-1$-index disclination is located at $X=-\frac{2\pi-\delta}{2\pi}r_{0}\cos(\delta/2)$ when this quantity is positive. As $\delta\rightarrow\pi^{+}$, the $-1$-index disclination coincides with the $+1$ disclination at the cone apex, and for smaller values of $\delta$ the grain boundary is free of point disclinations.  Note that this disclination dipole does not create a dislocation and is an example of a {\sl pincement}\cite{Klemanbook,Chen} that is so ``large'' as to have generated extra internal concentric layers, the dual to large Burgers vector dislocations \cite{Klemanbook}.

\section{Schlieren Textures}
\begin{figure}[t]
\begin{center}
\includegraphics[width=0.49\textwidth]{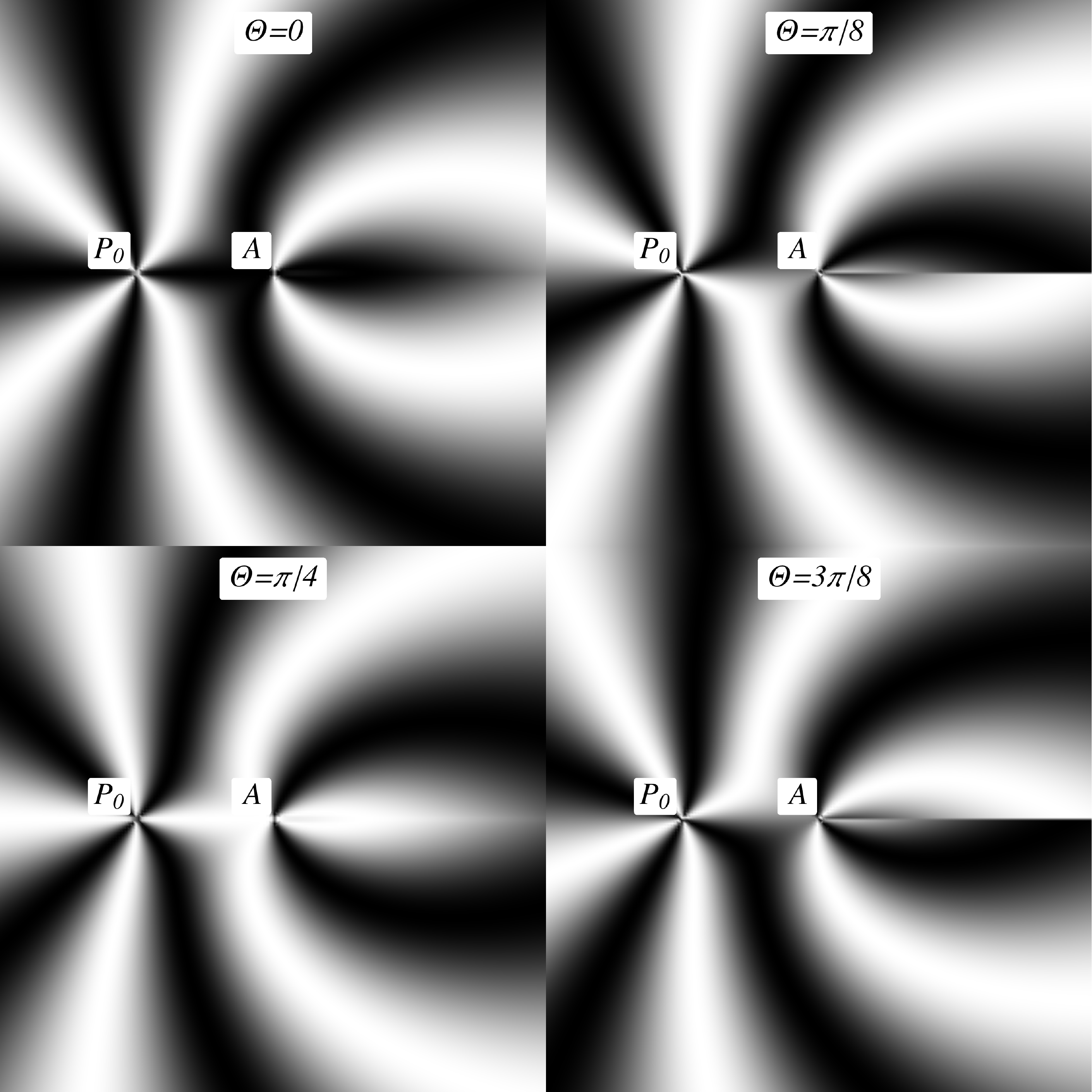}	
\end{center}
\caption{Schlieren textures for deficit angle $\delta=3\pi/2$, with the polarizer direction at angles of $\Theta=0$, $\pi/8$, $\pi/4$ and $3\pi/8$ with the $X$-axis, respectively. The analyzer direction rotates to remain perpendicular to the polarizer direction. $A$ and $P_0$ label the cone apex and the disclination location, respectively.}
\label{schlieren_rotate}
\end{figure}

How would these layer structures appear in an experiment? In examining nematic and smectic liquid crystalline textures, it is common to view the sample between a pair of perpendicularly crossed polarizers. The resulting schlieren texture, characterized by dark brushes on a bright background, reveals where in the sample the molecular orientation aligns on average with the direction of either polarizer. For a smectic-A liquid crystal on a conical substrate,  the molecules are normal to the layers aligned   along the unit vector field $\bn=(\bn_X,\bn_Y,\bn_Z)$ in the three-dimensional ambient space. If we were to view the sample between a pair of crossed polarizers parallel to the $\bx\by$ plane, we would measure $\bn_p$, the normalized horizontal projection of $\bn$. Denoting $\Theta$ as the oriented angle between the axis of one of the polarizers and the $\bx$ axis, it follows from \eq{proj_tang_vector}  that the intensity of the light observed at a point $(X,Y)$ is proportional to $\sin^2\left[2(\bphi-\bpsi-\Theta)\right]$.
Fig.~\ref{fig:layers_cone} shows the schlieren texture along with the layer structure for several choices of the deficit angle when $\Theta=0$. It is interesting to note that, besides the defect at $\bp_0$ (wavefront source), the schlieren texture also displays what is usually the signature of defects, the termination of dark brushes, at the apex and at another point farther down the cone at positive $X$. This occurs due to the grain boundary even for $\delta<\pi$, when the positive $X$ axis contains no topological defects. More surprising deviations from the usual rules governing schlieren textures are apparent when $\delta=3\pi/2$ and we rotate the polarizers, as shown in Fig.~\ref{schlieren_rotate}. To the right of the apex, dark brushes abruptly disappear into the horizontal axis from below, while other dark brushes spring into existence in the upper half-plane, as the polarizers turn counterclockwise. In an experiment, such a schlieren texture would be the clearest evidence of a grain boundary, demonstrating the range of ``missing" angles associated with a discontinuity in layer normals. Furthermore, the number of dark brushes emerging from the point defect at the apex is not constant and is odd for certain polarizer angles. In contrast, liquid crystalline textures that are continuous except (only) at point disclinations typically exhibit a constant, even number of dark brushes emanating from each disclination~\cite{hand}. This strange behavior can be understood by noting that the normalized horizontal projection $\bn_p$ of $\bn$ is not orthogonal to the projection of the layers on the $XY$ plane (as opposed to the 3D vector $\bn$ and the layers on the 3D cone which are, of course, orthogonal to each other). This can be easily seen in the left panels of Fig.~\ref{fig:layers_cone} and follows from the form of $ds^2$ in (\ref{eq:metric}). We will come back to this point in the next section when we discuss the conical bump.

\begin{figure}[hbt]
\begin{center}
\includegraphics[width=0.35\textwidth]{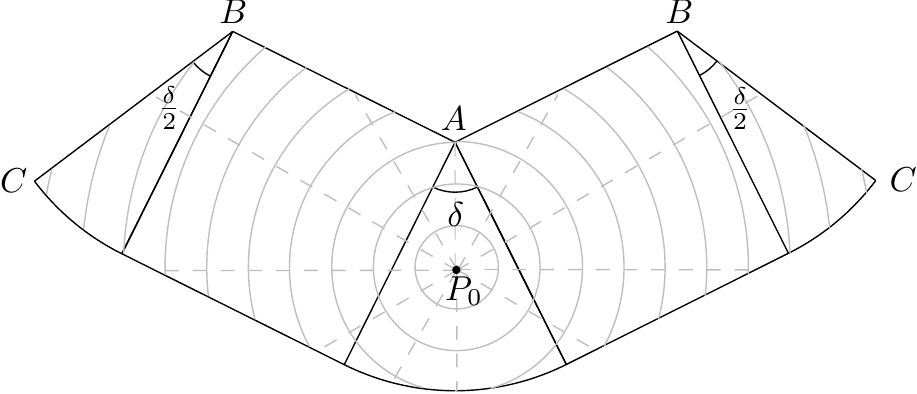} 
\includegraphics[width=0.24\textwidth]{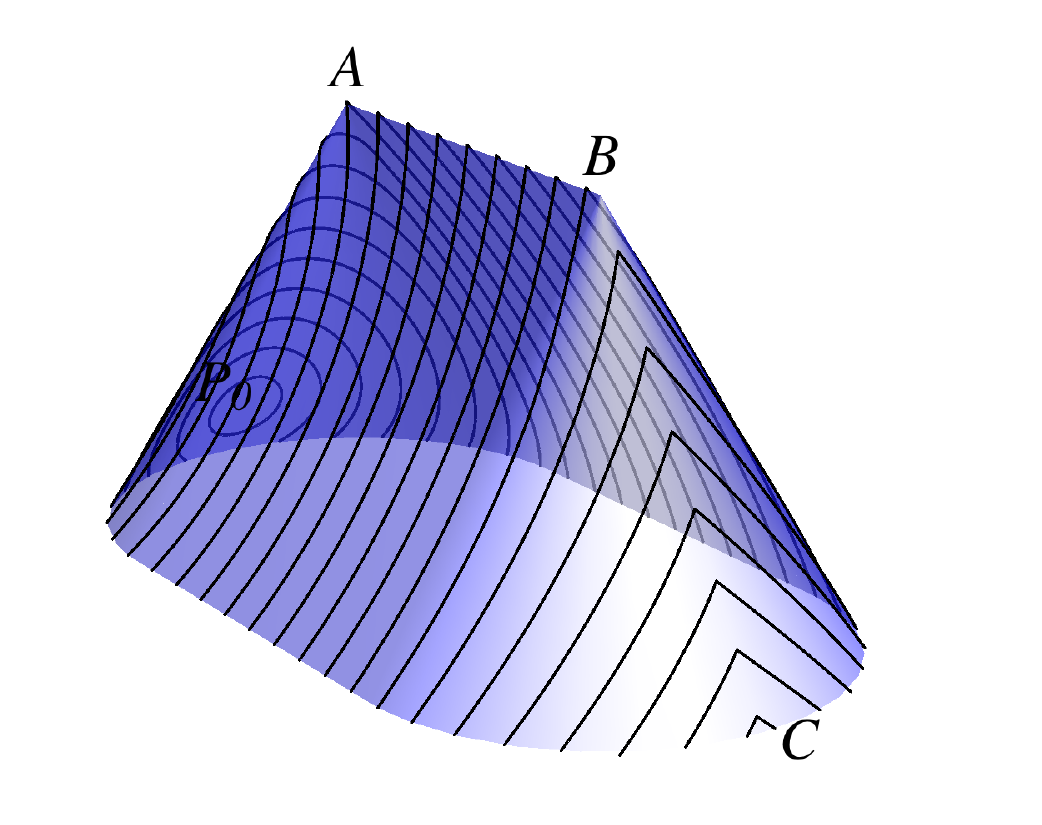} 
\includegraphics[width=0.23\textwidth]{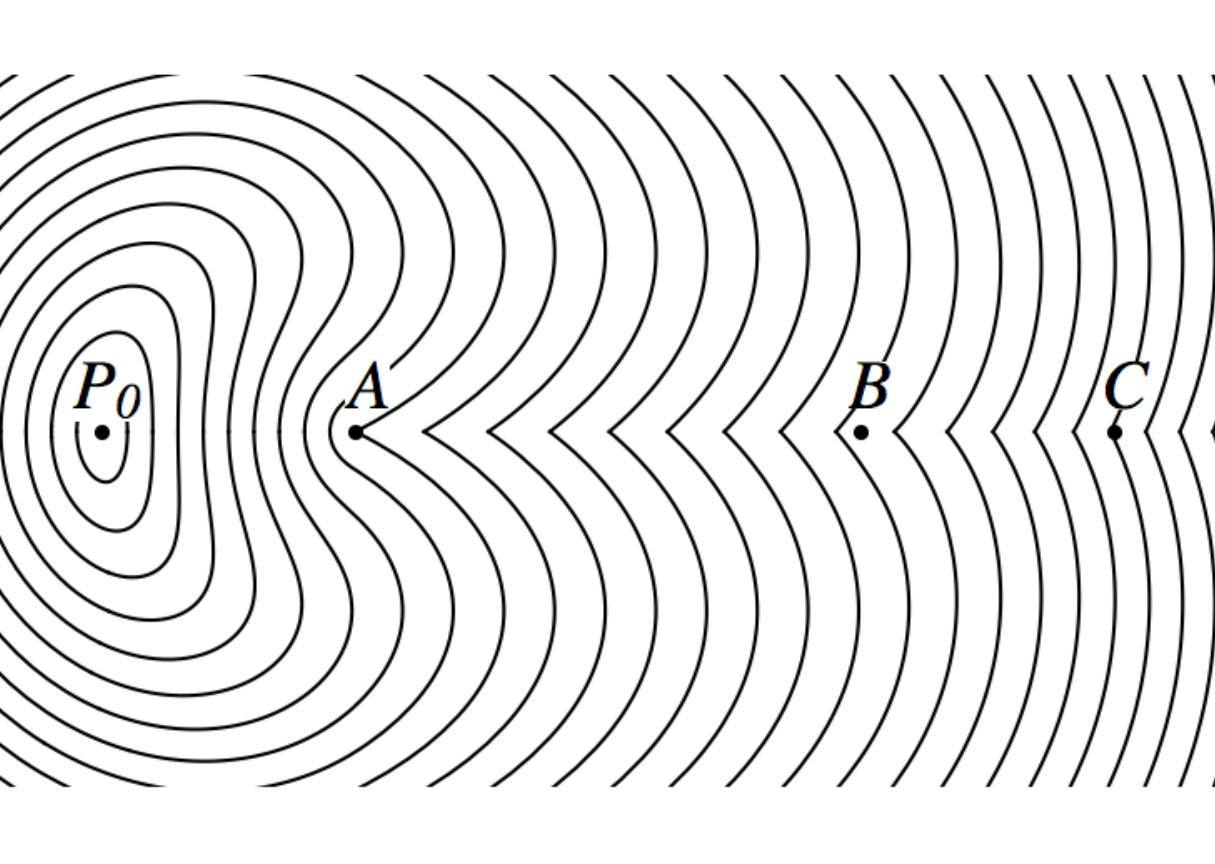} 
\end{center}
\caption{Substrate in the shape of a tent. The top panel shows a flattened model from which a tent can be constructed by gluing along the lines $AB$ and the lines $BC$. It also shows a point disclination at $P_0$ and its associated geodesics (dashed lines) and layers (solid lines). The second and third panels show the corresponding layers in the 3D tent and their 2D projection for the case when $\delta=\pi$ and $\overline{AB}=\overline{AP_{0}}=1$. $A$, $B$, and $P_0$ label the cones apices and the prescribed disclination location, respectively.}
\label{fig:tent}
\end{figure}

A slight generalization of the conical surface above is given by a tent, as shown in Fig~\ref{fig:tent} for $\delta=\pi$. When $\delta>\pi$, a $\pm1$-index disclination pair appears as on the cone, with the $+1$-index disclination located at the right endpoint of the tent ridge. The layer structure can be obtained identically as before by employing the flattened model shown in the first panel of Fig~\ref{fig:tent}.

\section{Smectics around edges}

The infinite cone has an isolated singularity at the apex. We have also considered the case of a tent, where the substrate has an edge, {\sl i.e.} a line where the surface is not smooth. Other examples are shown in the last three images of Fig.~\ref{fig:substrates}, for which a flattened model is not easily obtained because the edge is not straight in the $xy$-coordinate system. The geodesics of such surfaces will generally appear kinked at the edge. In fact, an argument similar to what is used in geometric optics, in connection to Fermat's principle, shows that a geodesic should cross an edge following Snell's law. This can be easily seen by noting that a geodesic is a curve with constant velocity that provides the path of minimal length --- and therefore minimal time --- between two given points. Note that the smectic analog of time is the number of layers through which the geodesic passes over a given distance.  Since the smectic layer spacing is the same on both sides of the interface, Snell's law implies that the angle of incidence equals the angle of ``refraction'' from the edge, where these angles are measured in the tangent planes on either side of the edge. The angle of refraction might differ from the angle of incidence if an interface separated two smectic phases of different layer spacing, as larger layer spacing is analogous to smaller index of refraction.  This could occur  in systems of immiscible smectics or at first-order transitions between different smectic phases of the same material.  But we digress.

\begin{figure}[htb]
\begin{center}
\includegraphics[width=0.19\textwidth]{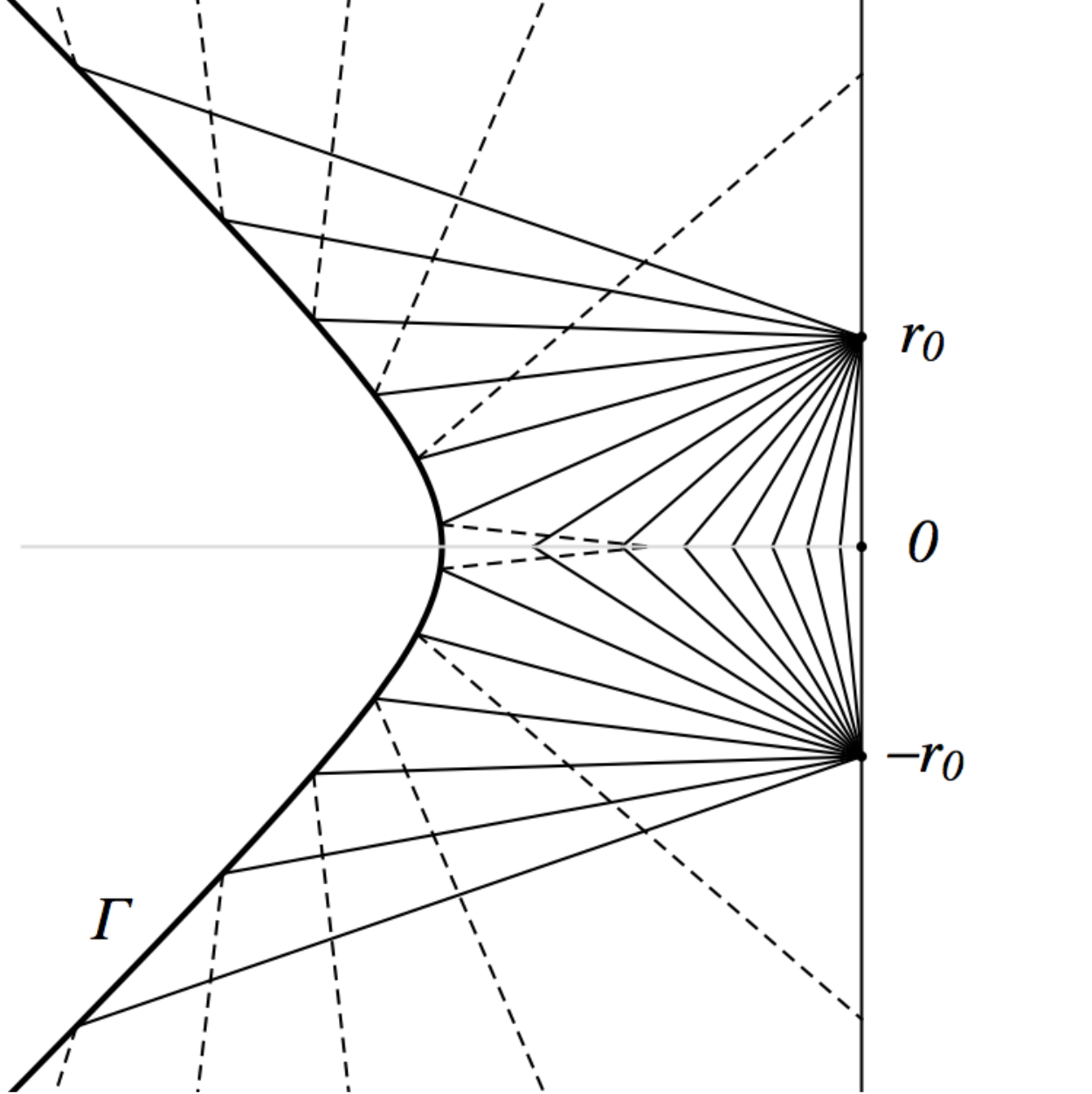}
\includegraphics[width=0.21\textwidth]{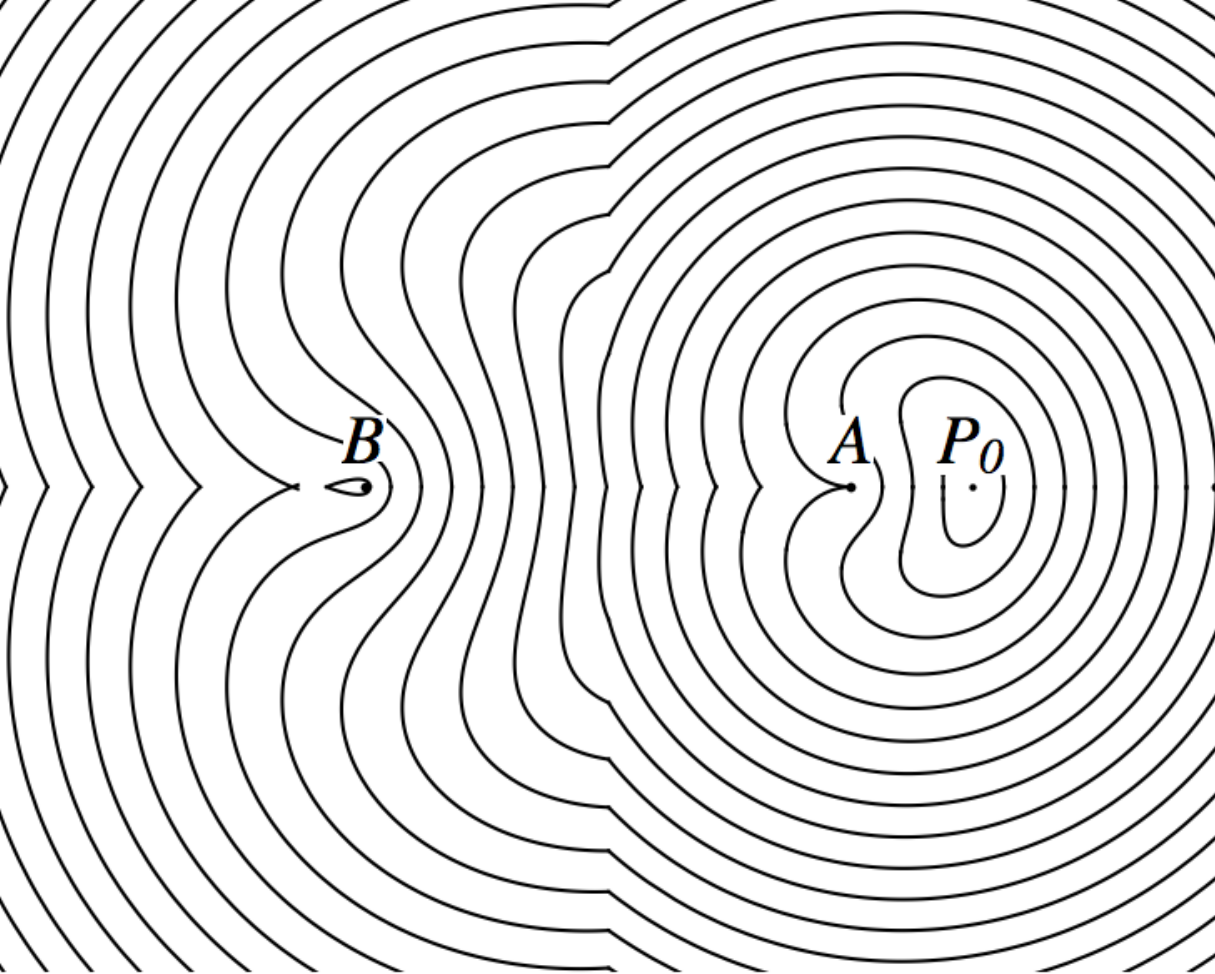} 
\includegraphics[width=0.49\textwidth]{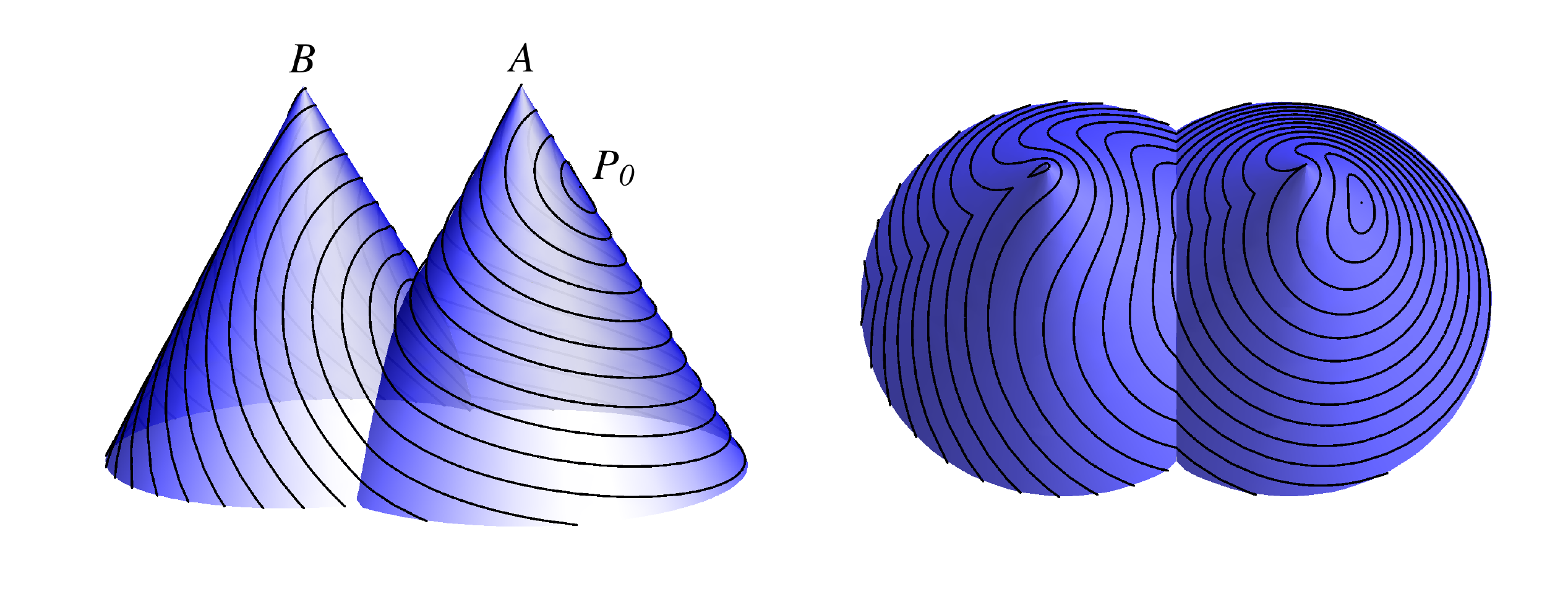} 
\end{center}
\caption{Smectic layers on two intersecting cones with deficit angle $\pi$ and apices $A$, $B$. The layer structure is determined by a point disclination at $P_{0}$, located a distance $r_{0}$ from apex $A$ as measured on the cone. The horizontal distance between the two apices is chosen to be $2 r_{0}$. The top left panel shows geodesics in the flattened out model of the rightmost cone. Notice that the cut line $L$ (Fig.~\ref{fig:flattened_cone}) coincides, in this case, with the positive and negative $y$ axes so that $(0,r_0)$ and $(0,-r_0)$ represent the same point. The intersection line is represented by $\Gamma$. Geodesics on the rightmost cone are represented by solid lines while the mirror reflection of those geodesics that enter the leftmost cone are depicted by dashed lines. The remaining panels show the layer structure on the cones.}
\label{fig:pass}
\end{figure}

Consider the ``crater'' on the upper right panel of Fig.~\ref{fig:substrates}. By symmetry, its layer structure (provided some boundary condition) can be immediately obtained from that on the single cone by reflection across an appropriate horizontal plane. A more interesting configuration is the mountain pass shown in the bottom left panel of Fig.~\ref{fig:substrates}. Here we can also use symmetry to simplify matters. Since the substrate has mirror symmetry across the vertical plane that contains the intersection, a geodesic that crosses the interface is simply the mirror image of a geodesic reflected through the edge. In the flattened model of the cone, the intersection will thus appear as a boundary $\Gamma$ that acts like a mirror, reflecting the ``incident rays'' according to the ``angle of incidence equals angle of reflection'' rule on the plane. This is displayed in Fig.~\ref{fig:pass}, which also shows the resulting layer structure for this case. 
Notice that a grain boundary is formed at the intersection between the surface and the $XZ$ plane for all points on the left of the rightmost apex, even between the cones.

Our discussion so far illustrates the general principle that, whenever curvature is present, the constraint of having equally spaced layers leads to singularities in their structure, with the appearance of cusps and grain boundaries. In particular, smectics on substrates composed of Gaussian bumps have been shown to provide an accessible system where these ideas take place~\cite{columns}. We now analyze a minimalist and localized version of the Gaussian bump,  the conical bump on the bottom right panel of Fig.~\ref{fig:substrates}.

\begin{figure}[htb]
\begin{center}
\includegraphics[width=0.49\textwidth]{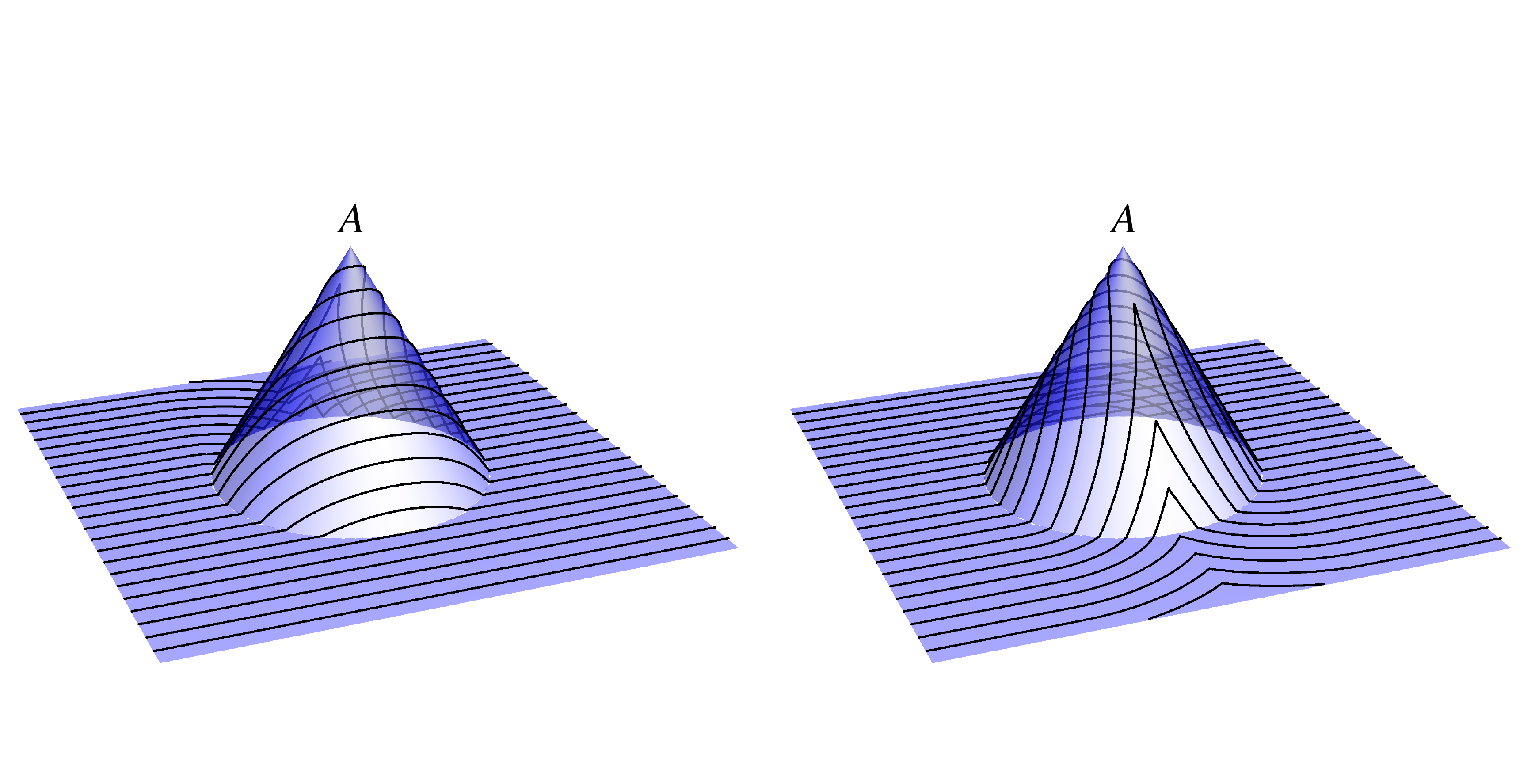}
\includegraphics[width=0.49\textwidth]{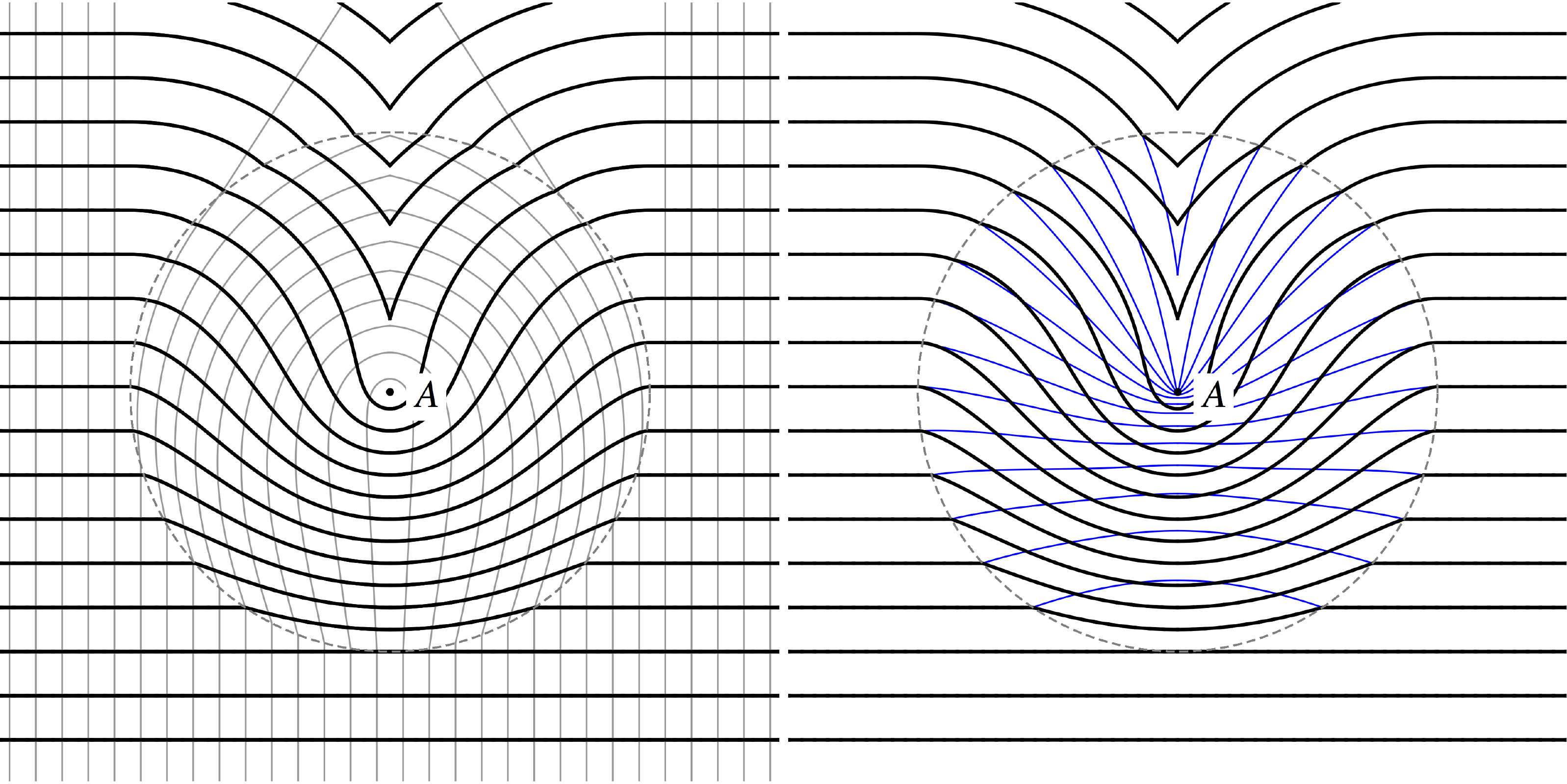}
\end{center}
\caption{Conical bump with deficit angle given by $\pi$. We consider as boundary conditions layers parallel to the $X$ axis at $Y\to-\infty$. The top panels show the layers on the 3D bump. The bottom left panel
shows the projected layers (black) and projected geodesics (gray) on the substrate; notice that they are not orthogonal to each other in the plane metric. The bottom right panel shows the projected layers (black) along with fictitious layers (thin blue lines) which are constructed by demanding orthogonality with the projected geodesics (see text). The intersection between the cone and the plane is depicted by a dashed circle and $A$ labels the cone apex.}
\label{fig:bump}
\end{figure}

We consider a situation with the boundary condition chosen to be layers parallel to the $X$ axis at $Y\to-\infty$. Before meeting the cone, a geodesic $\gamma$ which is normal to the layers is a straight line parallel to the $Y$ axis. At the interface it deflects according to Snell's law and then becomes a geodesic on the cone. There are two possibilities at this point, as the geodesics can become trapped in the cone or can escape. If $\gamma$ enters the cone near the $Y$ axis, i.e., with $X=X_0$ close to $0$, it will reach the $X=0$ plane before leaving the cone. As $X_0$ grows, $\gamma$ will leave the cone and become a straight line again before crossing the $X=0$ plane.  By symmetry, the same will happen to the geodesic corresponding to $-X_0$ and, as a result, a grain boundary will develop at the points of the cone located along the positive $Y$ axis.   An interesting observation can be made about the escaped geodesics. Since a cone is an axisymmetric surface, $\gamma$ must satisfy Clairaut's relation. This means that if $\rho(s)$ is the radial distance of the point $\gamma(s)$ (in the $XY$ plane) from the cone apex and $\alpha (s)$ is the angle that $\gamma'(s)$ makes with a longitude line of the surface, then $\rho(s)\sin \alpha(s)$ is constant for each such geodesic. At the boundary of the cone all the values of $\rho$ are the same and $\alpha$ is just the angle that $\gamma$ makes with the interface. Therefore, $\gamma$ enters and leaves the cone making the same angle with the edge. As before, to determine the trajectory of the geodesics inside the cone (and their exit point), we may either trace straight lines in the flattened model or solve the geodesic equations on the cone. This is all illustrated in Fig.~\ref{fig:bump} for a conical bump with deficit angle $\delta=\pi$.

We saw in the previous section that the projected layers of a non-planar surface are generally not orthogonal to their projected geodesics. This leads to schlieren textures exhibiting an odd behaviour  when the sample is analyzed with crossed polarizers parallel to the $XY$ plane.  Now, suppose that we do not initially know that the sample is really a curved surface and try to interpret it as a planar substrate. Apart from the fact that we would be surprised by the unusual pattern of brushes, we would also be led to identify a fictitious set of planar layers which are everywhere orthogonal to the projected geodesics, the latter being inferred from measurements under crossed polarizers. Since the projected geodesics are not geodesics on the plane by themselves, these fictitious layers cannot be equally spaced! This is illustrated by the bottom panels of Fig.~\ref{fig:bump}. Notice that, as expected, both the projected and the fictitious layers agree in the planar region but, inside the bump, the latter are highly compressed and have the opposite sign of curvature than the projected layers. This explains why the schlieren textures resulting from a non planar surface look so odd. If these fictitious layers were real, they would correspond to a high energy configuration, due to compression, and therefore would not represent the ground state. The transition to an equally spaced structure amounts precisely to escaping to the third dimension and assuming the layer configuration and shape of the bump.  The infinite strain in the fictitious layers near the cone apex signals this incompatibility as well.  Whether the fictitious layers and their geometry can be used as a surrogate to calculate the back reaction forces on a deformable surface is an open question.

\acknowledgements
It is a pleasure to acknowledge discussions with G.P. Alexander and B.G. Chen. RAM acknowledges financial support from FAPESP. DAB was supported by the NSF through a Graduate Research Fellowship. RAM and RDK thank the KITP for their hospitality while this work was completed and the support, in part, by NSF PHY11-25915. This work was supported in part by NSF Grant DMR05-47230 and a gift from L.J. Bernstein.

\end{document}